\documentstyle[12pt]{article}

\font\blackboard=msbm10 at 12pt
\font\blackboards=msbm7
\font\blackboardss=msbm5
\newfam\black
\textfont\black=\blackboard
\scriptfont\black=\blackboards
\scriptscriptfont\black=\blackboardss
\def\bb#1{{\fam\black\relax#1}}

\def\br{\bb R}

\newcommand{\ba}{\begin{array}}
\newcommand{\ea}{\end{array}}
\newcommand{\be}{\begin{equation}}
\newcommand{\ee}{\end{equation}}
\newcommand{\bea}{\begin{eqnarray}}
\newcommand{\eea}{\end{eqnarray}}
\newcommand{\beas}{\begin{eqnarray*}}
\newcommand{\eeas}{\end{eqnarray*}}

\font\cmss = cmss12

\def\laplace{{\kern1pt\vbox{\hrule height 1.2pt\hbox{\vrule width 1.2pt\hskip
  3pt\vbox{\vskip 6pt}\hskip 3pt\vrule width 0.6pt}\hrule height 0.6pt}
  \kern1pt}}
\def\scriptlap{{\kern1pt\vbox{\hrule height 0.8pt\hbox{\vrule width 0.8pt
  \hskip2pt\vbox{\vskip 4pt}\hskip 2pt\vrule width 0.4pt}\hrule height 0.4pt}
  \kern1pt}}

\def\roughly#1{\raise.3ex\hbox{$#1$\kern-.75em\lower1ex\hbox{$\sim$}}}

\def\IIB{\mbox{{\rlap{\cmss I} \hskip 0.7 true pt \hbox{\cmss IB}}}}

\def\p{\partial}

\newcommand{\NP}{{\em Nucl.\ Phys.\ }}
\newcommand{\PL}{{\em Phys.\ Lett.\ }}

\newcommand{\PRL}{{\em Phys.\ Rev.\ Lett.\ }}

\newcommand{\gone}[1]{}
\begin{document}
\pagestyle{plain}
\setcounter{page}{1}

\baselineskip16pt

\begin{titlepage}

\begin{flushright}
hep-th/9807225\\
PUPT-1806\\
RU-98-32\\
\end{flushright}
\vspace{16 mm}

\begin{center}
{\Large \bf Branes in the bulk of Anti-de Sitter space}

\vspace{3mm}

\end{center}

\vspace{8 mm}

\begin{center}
\setcounter{footnote}{2}

Michael R.\ Douglas${}^1$ and Washington Taylor
IV${}^{2,}$\footnote{Address after August 1998: Center for Theoretical
Physics, Massachusetts Institute of Technology, Cambridge MA 02139,
{\tt wati@mit.edu}}

\vspace{3mm}

${}^1${\small \sl Department of Physics and Astronomy} \\
{\small \sl Rutgers University} \\
{\small \sl Piscataway, New Jersey 08555, U.S.A.} \\
{\small and} \\
{\small \sl I.H.E.S., Le Bois-Marie, Bures-sur-Yvette, 91440 France} \\  
{\small \tt mrd@physics.rutgers.edu}

\vspace{3mm}

${}^2${\small \sl Department of Physics} \\
{\small \sl Joseph Henry Laboratories} \\
{\small \sl Princeton University} \\
{\small \sl Princeton, New Jersey 08544, U.S.A.} \\
{\small \tt wati@princeton.edu}

\end{center}

\vspace{8mm}

\begin{abstract}

We consider (3 + 1)-dimensional ${\cal N} = 4$ super Yang-Mills theory
with a nonvanishing scalar Higgs vacuum expectation value, and compare
this theory to AdS supergravity with branes in the bulk.  We show that
the one-loop effective potential for excitations of the Yang-Mills
field agrees with the classical linearized potential for brane waves
in the AdS picture in the limit of long wavelengths.
This supports the idea that the AdS/CFT correspondence fits into
string theory as expected from previous work.

\end{abstract}

\vspace{8mm}
\begin{flushleft}
July 1998
\end{flushleft}
\end{titlepage}
\newpage

In \cite{Maldacena-AdS}, it was conjectured by Maldacena that there is a
remarkable equivalence between ${\cal N} = 4$ super Yang-Mills (SYM) theory
in (3 + 1) dimensions and type \IIB\  string theory on $AdS_5 \times
S^5$.  This conjecture has been sharpened and tested in many ways.
In the approaches of Gubser, Klebanov and Polyakov
\cite{gkp-2} and of Witten \cite{Witten-AdS1} the connection between SYM
theory and AdS string theory is made precise by placing the 3-branes of the
SYM theory on the boundary of the AdS space, and associating
correlation functions of operators in the boundary SYM theory with
supergravity correlation functions subject to certain boundary
constraints.  In this picture, it is assumed that the adjoint Higgs
fields in the SYM theory vanish so that the theory is at a conformal
point.  

One way to motivate the conjecture is to consider a system of a large
number $N$ of coincident D-branes in the string theory.  On the one
hand, at substringy distances from the branes the dynamics of the
system is well-described by truncating to the gauge theory on the
branes \cite{DKPS}.  On the other hand, for $g_s N \gg 1$ the
curvature at these distances is small in string units.  This suggests
that questions which can be formulated in gravitational terms (e.g.,
{\em what is the metric?})  are controlled by supergravity with small
stringy corrections, as pointed out in \cite{DPS}.

According to this argument, the AdS/SYM correspondence is a particular
limit of the dynamics of branes in string theory, and we can consider
other configurations of the branes before taking the limit.  In
particular, we can consider the Coulomb branch of the gauge theory,
i.e. nonzero vacuum expectation values for the scalar fields,
corresponding to branes in the bulk of the AdS space.  This
possibility was also suggested in \cite{Maldacena-AdS} but it has not
played a significant part in the developments so far; in fact, the
general belief at this point seems to be that the SYM/AdS connection
only makes sense when the scalar vevs vanish.

In this letter we exhibit an example in which this possibility is realized --
an SYM configuration with nonvanishing scalar expectation values
corresponds to AdS supergravity with branes in the bulk.

We will work with large $N$
(3 + 1)-dimensional SYM theory in a supersymmetric vacuum in which
the adjoint Higgs fields take the form
\[
X_i = \left(\begin{array}{ccc}
 x_i & 0 & 0\\
0 & \tilde{x}_i& 0\\
0 & 0 & {\bf 0}
\end{array}\right)
\]
Here ${\bf 0}$ indicates the vev for $N-2$ eigenvalues of the Higgs
fields, while $x_i$ and $\tilde x_i$ are single eigenvalues.

Our basic assertion will be that the eigenvalues $x_i$ and $\tilde x_i$
are the positions of $3$-branes in
the usual way, both before and after taking the scaling limit of
\cite{Maldacena-AdS}.  The simplest argument for this is ``what else
can it be?'' -- assuming the conjecture, the probe $3$-branes must exist
in the theory, and there is no other candidate description.
However let us see to what extent we can find evidence for this statement
without assuming the result.

By using the $O(6)$
symmetry of the theory we can rotate the fields into the form
\begin{equation}
X_1 = \left(\begin{array}{ccc}
 r & 0& 0\\
0 & \tilde{r} \cos \theta& 0 \\
0& 0& {\bf 0}
\end{array}\right) \;\;\;\;\;
X_2 = \left(\begin{array}{ccc}
 0 & 0& 0\\
0 &\tilde{r}\sin \theta& 0\\
0& 0& {\bf 0}
\end{array}\right) 
\label{eq:Higgs}
\end{equation}
where $ r, \tilde{r}$ are the magnitudes of the 6-vectors
$x, \tilde{x}$ and $\theta$ is the angle between these vectors.

We now consider the fields (\ref{eq:Higgs}) as a
background and add fluctuations of the gauge field on the two
branes.  Denoting the $U(1)$ field strength on the first (second) brane by
$\widehat{F}$ ($\widetilde{F}$) respectively, the one-loop effective action
describing the leading interaction between the two fields is proportional to
\cite{1001,Vipul-rvu}
\begin{eqnarray*}
& &\int \prod_{i} d^4 p_i 
\left[F^\mu{}_\nu (p_1) F^\nu{}_\lambda (p_2) F^\lambda{}_\kappa (p_3)
F^{\kappa}{}_\mu (p_4) 
-\frac{1}{4}  (F^\mu{}_\nu (p_1) F^\nu{}_\mu (p_2))(F^\kappa{}_\lambda (p_3)
F^\lambda{}_\kappa (p_4))\right] \\
& &\hspace{0.7in}
\times \delta^4 (p_1 + p_2 + p_3 + p_4)
\left[G (p_1, p_2, p_3, p_4) + {\rm permutations} \right]
\end{eqnarray*}
where $F = \widehat{F}-\widetilde{F}$, 
and
\begin{equation}
G (p_1, p_2, p_3, p_4) = \int d^4 k
\frac{1}{ ( \rho^2 + k^2) (\rho^2 +(k-p_1)^2) (\rho^2 + (k-p_1-p_2)^2)
(\rho^2 + (k + p_4)^2)} 
\label{eq:momentum}
\end{equation}
with $\rho = \sqrt{(x_i-\tilde{x}_i)^2}$.  
If we  have $N = 2$,  this one-loop result determines the
four-derivative terms in the effective action exactly,
by the non-renormalization theorem of \cite{Dine-Seiberg}.

In general, at this order the result is a sum of pairwise interactions
between the branes.  Although this contribution to the interaction
between pairs of branes at ${\bf 0}$ is infrared divergent, the
interaction between the two branes at $x_i$ and $\tilde x_i$ is not
affected by this.  We can make a stronger statement if we consider the
limit from the Coulomb branch, i.e. taking the $N-2$ eigenvalues
to zero, because away from zero the non-renormalization theorem still
holds.  Since this is uniform as we take the limit and there is no
remaining dimensional parameter to control corrections, it is very
plausible that the result remains true in the limit.  We will see
evidence for this shortly.

This result determines the leading scattering of ``brane
waves,'' small fluctuations of the massless modes on branes one and two.
The essential statement we are making at this point is that such modes
exist and are created by operators such as $\hat F$ and $\tilde F$,
and are weakly interacting.  This is clearly true in
the original brane picture and at small $g_s N$ but one might argue
the point in gauge theory at large $g_s N$.
A gauge theory justification for our statement
is that we take $g_s$ small (even when $g_s N$ is large),
while the interactions to the conformal sector of the system are controlled
by the masses $r$ and $\tilde r$.  This is not to say that such interactions
are negligible but that they can produce the dynamics of branes in a
\IIB\ string theory background.

Given this statement, it is clear that the result is an observable
scattering amplitude.
We are now interested in comparing it with expectations from
supergravity; i.e. we consider the two brane waves as sources of
stress-energy (and the other supergravity fields) and compute their
interactions from this point of view.

We first consider the zero-momentum limit; i.e.
the case where the fluctuations of the gauge
field are constant on both branes.  In this case, there is an
effective potential between the branes of the form
\be
\frac{1}{\rho^4} 
\left[F^\mu{}_\nu F^\nu{}_\lambda F^\lambda{}_\kappa
F^{\kappa}{}_\mu 
-\frac{1}{4}  (F^\mu{}_\nu F^\nu{}_\mu)(F^\kappa{}_\lambda
F^\lambda{}_\kappa)\right] 
\label{eq:effpotzero}
\ee
By explicitly expanding this expression in terms of the field
strengths $\widehat{F}$ and $\widetilde{F}$ on the two branes, and grouping
the terms according to their Lorentz index structure, the interaction
between the two branes can be expressed  in terms of a sum of current-current
interactions for the stress tensor as well as the sources for the R-R
and other NS-NS fields in the \IIB\  theory.  An analogous decomposition
was given explicitly in the context of Matrix theory and
11-dimensional supergravity in \cite{Dan-Wati2}.  In the \IIB\ theory
the identification of currents is slightly different from that given
in \cite{Dan-Wati2}, although the essential structure is the same; in
this case, all the current-current interactions carry an overall
factor of $1/\rho^4$.  As a simple example of the types of interactions which
appear in this decomposition,
there are terms of the form
\[
\frac{1}{\rho^4}  \left[
\left(\widehat{F}^{\mu \nu} \widehat{F}_{\nu \mu} \right)
\left(\widetilde{F}^{\lambda \rho} \widetilde{F}_{\rho \lambda} \right)\right]
\;\;\;\;\; {\rm and}\;\;\;\;\;
\frac{1}{\rho^4}  \left[ \left(
\widehat{F}^{\mu \nu} \widehat{F}_{\nu \lambda} \right)
\left(\widetilde{F}^{\lambda \rho} \widetilde{F}_{\rho \mu} \right)\right].
\]
These terms correspond to pieces of the effective potential arising
from dilaton and graviton exchange
\cite{Klebanov-absorption,gkt}.  
Another example is the term of the form 
\[
\frac{1}{\rho^4}  \left[
\left(\widehat{F}^{\mu \nu} \widehat{F}_{\nu \lambda}
\widehat{F}^{\lambda \rho} \right)  \left(\widetilde{F}_{\rho \mu} \right) \right].
\]
As shown in \cite{Das-Trivedi,Ferrara}, the cubic
expression in terms of $\widehat{F}$ is the leading operator which couples
to the field $B_{\rho \mu}$, for which $\widetilde{F}_{\rho \mu}$ acts as
the source.  Although we are only considering fluctuations of the
gauge fields on a pair of single 3-branes, it is straightforward to
generalize this discussion to include non-abelian scalar vev's in a
pair of 3-brane clusters, which are necessary to describe fields with
transverse polarization.
For such configurations there are ``higher moment'' contributions to
(\ref{eq:effpotzero}) of the form $F^4 X^n/\rho^{4 +n}$, as discussed
in \cite{Dan-Wati2}.  These terms should describe the interactions between
the higher partial waves on $S^5$ discussed, for example, in
\cite{Klebanov-absorption}.

For $N = 2$, the SYM theory is just the low energy world-volume
dynamics of 3-branes in flat space.
The first and second 3-branes are separated by a distance
\[
\rho = \sqrt{\sum_{i} (x_i-\tilde{x}_i)} 
= \sqrt{r^2 + (\tilde{r})^2 -2r\tilde{r} \cos \theta}.
\]
{}From this point of view, the function
$1/\rho^4$ appears because it is the Green's function for the
flat-space Laplacian in the six dimensions transverse to the 3-brane:
\[
{\Delta}_{{\bb R}^6} = -{1\over r^5}\p_r r^5 \p_r - {1\over r^2} \Delta_{S^5}
\]
\[
{\Delta}_{{\bb R}^6} \frac{1}{\rho^4}  \sim
\delta^6 ( x-\tilde{x}).
\]
We thus see the usual one-loop agreement between gauge theory and
supergravity.

At large $N$ and $g_s N$, and in the scaling limit
$r \sim l_s^2 \rightarrow 0$,
the AdS interpretation is appropriate.  
The two probe branes are placed in the
$AdS_5 \times S^5$ metric, the scaling limit
of the metric around the $N-2$ branes:
\begin{equation}
 ds^2 = r^2(-dt^2 + dx_{\parallel}^2) + {dr^2\over r^2} + d \Omega_5^2.
\label{eq:AdS-metric}
\end{equation}
The leading long-distance gravitational interaction between a pair of
3-branes with excited gauge or transverse fields
is described (in an appropriate gauge) by a propagator ${\cal P}
(x; \tilde{x})$ which is the
Green's function for the Laplacian in the metric (\ref{eq:AdS-metric}).  
This is a priori different from the flat space prediction.   

In the case where the excitations
on the 3-branes are constant in the parallel directions $x_\parallel$,
the propagator must satisfy
\begin{equation}
\frac{1}{ \sqrt{-g}}  \partial_\mu \sqrt{-g} g^{\mu \nu}
\partial_\nu{\cal P} (r, \omega; \tilde{r},
\tilde{\omega})
\sim \delta^6 ( r-\tilde{r}, \omega -\tilde{\omega})
\label{eq:AdS-propagator}
\end{equation}
where $(r,\omega) \in{\bb R} \times S^5$.
In fact, the flat-space Green's function $1/\rho^4$ 
turns out to solve (\ref{eq:AdS-propagator}).
This can be seen most directly by working out the $AdS_5\times S^5$
Laplacian.  On functions depending only on the coordinates $(r,
\omega)$ this Laplacian acts as
\beas
\Delta_{AdS} & =& -\frac{1}{r^3} \p_r r^5 \p_r - \Delta_{S^5} \\
&=&  r^2\Delta_{{\bb R}^6} .
\eeas
Thus the effective potential (\ref{eq:effpotzero}) agrees with expectations
from AdS supergravity as well.  

Agreement in both limits suggests that the result will agree at arbitrary
$g_s N$.  This is also true and is the supergravity
counterpart of the nonrenormalization conjecture (the generalization
we suggested of \cite{Dine-Seiberg} to the conformal limit).
Before rescaling, the metric around $N$ $3$-branes is
\[
ds^2 = f(r)^{-1/2} (-dt^2 + dx_{\parallel}^2) 
+ f(r)^{1/2} (dr^2 + r^2 d \Omega_5^2 )
\label{eq:brane-metric}
\]
with
\[
f(r) = 1 + {g_s N\over r^4} .
\]
One can again check that $\Delta_{general} = f^{-1/2} \Delta_{\bb R^6}$
allowing the same Green's function to solve this Laplacian as well.

We now briefly discuss the case of non-zero momentum.
One can check that turning on non-zero momenta in
(\ref{eq:momentum}) leads to corrections controlled by the scale $l_s$,
for which we have not found a supergravity interpretation.  
What is clear from the expression (\ref{eq:momentum}) is
that the result is analytic in the external momenta at $p=0$,
and has an expansion in $p/r$, as usual in effective field theory.
In other words, a derivative $d/dp$ can be expressed in terms of
the derivative $d/dr$ (and less singular terms as $r\rightarrow 0$).
As we discuss below, we interpret this in terms of an uncertainty
relation for the position of the brane.

\medskip

In this note we have observed that the leading interaction between
brane waves on $3$-branes agrees when computed as a quantum effect in
gauge theory, and as an interaction mediated by bulk fields in
supergravity, both in a flat background and in the AdS background
corresponding to a large number of additional $3$-branes.

Besides providing a further example of an agreement between super
Yang-Mills theory and supergravity, we consider the 
result to be significant evidence for the 
following identification: in a supersymmetric vacuum in the gauge theory,
a non-zero eigenvalue for the scalar vev coresponds to the position of 
an {\it individual} $3$-brane.
This is the naive correspondence suggested by the brane picture, but
it is worth stating it again in the context of the AdS/CFT correspondence.

Note that this does not necessarily mean that we should consider the
$N-2$ branes in our illustration as located at the point $x=0$.
Conceptually, if we are regarding supergravity as coming out of 
large $N$ gauge theory dynamics, the simplest picture (as has been
described by Maldacena; see also \cite{BDHM})
is not to consider the conformal sector of 
the theory as localized anywhere.
Excitations on the $N-2$ branes are not free brane waves --
once we consider $g_s(N-2)$ large
there is no control over infrared divergences in our computation, and no
justification for the result (\ref{eq:effpotzero}).

Another class of vacua mentioned in \cite{Maldacena-AdS}, and
motivated from the brane point of view is to split the branes into
groups with $N = \sum N_i$ each with a different scalar vev.
If we take the scaling limit of these vevs, the natural conjecture 
is that when all $g_s N_i$ become large, this corresponds to supergravity
in the background of a multicenter solution.  Again one is not asserting
that ``the $N_i$ branes are located at these points''
because their excitations are
not brane waves; however a probe brane could be localized. 
This example has been studied in \cite{MinWar}; many questions remain.
We mention that our result also works in this case.

The identification also gives us a class of local observables
in AdS space, the operators which create excitations on the probe branes.
An important point however is that the localization of the
probe branes appears to be precise only in the zero longitudinal
momentum limit.
In work such as \cite{BDS,DKPS} it was observed that to localize the
branes in transverse space, one needed arbitrarily low energies,
because the energy of the states responsible for the interactions
went as $E \sim r/l_s^2$.
This can also be formulated as an uncertainty principle \cite{Yoneya}:
using $\Delta E \Delta t > 1$ we have $\Delta r \Delta t > l_s^2$ and
we need arbitrarily large time to make such a measurement.
The scaling limit of \cite{Maldacena-AdS} takes 
$r \sim l_s^2 \rightarrow 0$, so the uncertainty principle survives,
as observed in \cite{Li-Yoneya}.

We interpret the corrections to the interaction of brane waves at
nonzero momentum (compared to supergravity)
as a manifestation of this uncertainty principle.
Given an uncertainty $\Delta p$ around zero in energy-momentum, we saw that
the difference between the gauge theory and supergravity potentials $V$ can be
approximated by our observations as $\Delta p~d/dr~V$. 
This is what would come from
an uncertainty $\Delta r~d/dr~V$ with $\Delta r \sim \Delta p \sim \Delta E$
as above.  It will be interesting to follow this up and especially to
understand if this is a universal constraint on our ability to formulate
local observables in the AdS language.

Finally, this paper only discussed the case of gauge theory
on $\br^{3,1}$, while the AdS/SYM correspondence can be formulated for
more general base spaces, for example $S^3 \times \br$.
One may ask: how does one describe a probe $3$-brane in this case?
Doesn't the lifting of the moduli space described in \cite{Witten-AdS1}
eliminate the configurations we described here?

The answer is clear once one takes the conformal symmetry of the theory
into account.  In some cases one can relate the theory on different
base spaces by
conformal transformations, and map the simple supersymmetric vacua we
describe here into solutions on a different base space.
More generally, one needs to find solutions which preserve $16$ of
the superconformal symmetries; we predict that appropriate candidates
will exist.  
All of these are
time-dependent solutions, so there is no contradiction with the
nonexistence of a moduli space.

Applying conformal transformations allows us to produce new solutions
in $\br^{3,1}$ as well.  Since the corresponding AdS isometries 
can map points
within the coordinate patch (\ref{eq:AdS-metric}) to points outside
\cite{HoroOoguri}, these describe new
$3$-branes extending across the asymptotic observer's horizon.
These will be studied in future work.

\section*{Acknowledgements}

This work was carried out at the workshop on Dualities in String
Theory at the ITP Santa Barbara, and we would like to thank David Gross,
Dan Hone and the entire staff for providing a fantastic environment
for doing physics.

The work of MRD is supported in part by DOE grant
DE-FG02-96ER40559.  The work of WT is
supported in part by the National Science Foundation (NSF) under
contract PHY96-00258.

We also call the reader's attention to \cite{Berkooz}, which
uses $3$-branes to study microscopic aspects of the AdS/SYM correspondence.


\begin{thebibliography}{10}

\bibitem{Maldacena-AdS}
J.\ Maldacena, ``The large N limit of superconformal field theories and
  supergravity,'' {\tt hep-th/9711200}.

\bibitem{gkp-2}
S.\ S.\ Gubser, I.\ R.\ Klebanov and A.\ M.\ Polyakov, ``Gauge theory
  correlators from non-critical string theory,'' {\tt hep-th/9802109}.

\bibitem{Witten-AdS1}
E.\ Witten, ``Anti-de Sitter space and holography,'' {\tt hep-th/9802150}.

\bibitem{DKPS}
M.\ R.\ Douglas, D.\ Kabat, P.\ Pouliot, and S.\ Shenker, ``D-Branes and Short
  Distances in String Theory,'' \NP {\bf B485} (1997) 85, {\tt hep-th/9608024}.

\bibitem{DPS}
M.\ R.\ Douglas, J.\ Polchinski and A.\ Strominger, {\em JHEP} {\bf 12} (1997)
  003, {\tt hep-th/9703031}.

\bibitem{1001}
S.\ J.\ Gates Jr., M.\ T.\ Grisaru, M.\ Rocek and W.\ Siegel, {\em SUPERSPACE}
  or {\em One thousand and one lessons in supersymmetry}, (Benjamin/Cummings
  Publishing Company, Reading, 1987).

\bibitem{Vipul-rvu}
V.\ Periwal and R.\ von Unge, ``Accelerating D-branes,'' {\tt hep-th/9801121}.

\bibitem{Dine-Seiberg}
M.\ Dine and N.\ Seiberg, ``Comments on higher derivative operators in some
  SUSY field theories,'' {\tt hep-th/9705057}.

\bibitem{Dan-Wati2}
D.\ Kabat and W.\ Taylor, ``Linearized supergravity from Matrix theory,'' \PL
  {\bf B426} (1998) 297-305, {\tt hep-th/9712185}.

\bibitem{Klebanov-absorption}
I.\ R.\ Klebanov, \NP {\bf B496} (1997) 231, {\tt hep-th/9702076}.

\bibitem{gkt}
S.\ S.\ Gubser, I.\ R.\ Klebanov and A.\ A.\ Tseytlin, \NP {\bf B499} (1997)
  217, {\tt hep-th/9703040}.

\bibitem{Das-Trivedi}
S.\ R.\ Das and S.\ P.\ Trivedi, ``Three-brane action and the correspondence
  between ${\cal N} = 4$ Yang-Mills theory and anti-de Sitter space,'' {\tt
  hep-th/9804149}.

\bibitem{Ferrara}
S.\ Ferrara, M.\ A.\ Lledo and A.\ Zaffaroni,
``Born-Infeld Corrections to D3 brane Action in $AdS_5\times S_5$
and N=4, d=4 Primary Superfields,''
{\tt hep-th/9805082}.

\bibitem{BDHM}
T.\ Banks, M.\ R.\ Douglas, G.\ Horowitz and E.\ Martinec, to appear.

\bibitem{Yoneya}
T.\ Yoneya, {\em Mod.\ Phys.\ Lett.} {\bf A4} (1989) 1587; M.\ Li and T.\
  Yoneya, \PRL {\bf 78} (1997) 1219, {\tt hep-th/9611072}.

\bibitem{Li-Yoneya}
M.\ Li and T.\ Yoneya, ``Short-distance space-time structure and black holes in
  string theory: a short review of the current status,'' {\tt hep-th/9806240}.

\bibitem{BDS}
A.\ Sen, \NP {\bf 475} (1996) 562, {\tt hep-th/9605150};
T.\ Banks, M.\ R.\ Douglas, N.\ Seiberg, 
	\PL {\bf 387} (1996) 278, {\tt hep-th/9605199};
N.\ Seiberg, \PL {\bf 384} (1996) 81, {\tt hep-th/9606017}.

\bibitem{MinWar}
J.\ Minahan and N.\ Warner, 
``Quark Potentials in the Higgs Phase of Large N Supersymmetric
Yang-Mills Theories,'' {\tt hep-th/9805104}.

\bibitem{HoroOoguri}
G.\ Horowitz and H.\ Ooguri, \PRL {\bf 80} (1998) 4116-41180, {\tt
hep-th/9802116}.

\bibitem{Berkooz}
M. Berkooz, {\tt hep-th/9807230}.

\end{thebibliography}
\bibliographystyle{plain}

\end{document}